\documentclass[twocolumn,showpacs,preprintnumbers,amsmath,amssymb]{revtex4}

\usepackage{graphicx,amssymb,lineno}

\usepackage{dcolumn}
\usepackage{bm}

\newcommand{\fecosi}[0]{Fe$_{1-x}$Co$_{x}$Si }

\newcommand{\mncosi}[0]{Mn$_{1-x}$Co$_{x}$Si }

\newcommand{\mnfesi}[0]{Mn$_{1-x}$Fe$_{x}$Si }
\begin{document}

\title{Spin Glass ground state in \mncosi}

\author{J. Teyssier}
\author{E. Giannini}
\author{V. Guritanu}
\author{R. Viennois}
\author{D. van der Marel}

\affiliation{D\'epartement de Physique de la Mati\`ere Condens\'ee, Universit\'e de Gen\`eve, Quai Ernest-Ansermet 24, 1211 Gen\`eve 4, Switzerland}

\author{A. Amato}
\affiliation{Laboratory for Muon-Spin Spectroscopy, Paul-Scherrer Institut, 5232 Villigen PSI, Switzerland}

\author{S. N. Gvasaliya}
\affiliation{Laboratory for Neutron Scattering, ETHZ $\&$ Paul-Scherrer Institut, 5232 Villigen PSI, Switzerland}
\date{\today}

\begin{abstract}
We report the discovery of a new spin glass ground state in the transition metal monosilicides with the B20 crystallographic structure. Magnetic, transport, neutron and muon investigation of the solid solution \mncosi have revealed a new dome in the phase diagram with evidence of antiferromagnetic interactions. For Mn rich compounds, a sharp decrease of the Curie temperature is observed upon Co doping and neutron elastic scattering shows that helimagnetic order of MnSi persists up to $x=0.05$ with a shortening of the helix period. For higher Co ($0.05<x<0.90$) concentrations, the Curie-Weiss temperature changes sign and the system enters a spin glass state upon cooling ($T_g=9$~K for $x_{Co}=0.50$), due to chemical disorder. In this doping range, a minimum appears in the resistivity, attributed to scattering of conduction electron by localized magnetic moments.
\end{abstract}

\maketitle

\section{Introduction}

The transition metal monosilicides TM-Si with the B20 cubic structure (TM=Cr,Mn,Fe,Co) are the object of intense studies due to their interesting and various magnetic ground states. When increasing the number of electrons of the transition metal one crosses CrSi, a Pauli paramagnet \cite{Shinoda1966,Wernick1972}, MnSi, an itinerant helimagnetic metal \cite{grigoriev2006}, FeSi, a paramagnetic insulator \cite{Paschen1997}, and CoSi, a diamagnetic metal \cite{Wernick1972,Han2001,Nakanishi1980}. The B20 structure remains stable up Co$_{0.65}$Ni$_{0.35}$Si \cite{Teyssier2008}. NiSi is a diamagnetic metal that crystallizes in the B31 orthorhombic structure \cite{Wernick1972,Meyer1997}.

If the binary B20 family is already a rich catalog of various electronic states, it is made even wider by mixing TM atoms to form ternary solid solutions. Fe$_x$Co$_{(1-x)}$Si exhibits itinerant helimagnetic metallic behavior like MnSi for $0.4<x<0.9$ ($T_c=60K$ for $x=0.6$) although the two end-compounds FeSi and CoSi do not exhibit any magnetic order \cite{Moriya1973,Manyala2000,Manyala2004,Onose2005}. Doping FeSi with Mn has revealed that the unscreened Kondo effect was at the origin of non-Fermi liquid behavior \cite{Manyala2008}. In this material, the Curie-Weiss temperature becomes negative when less than 80\% of Fe is replaced by Mn. Despite of this evidence of an antiferromagnetic exchange, no ordering is reported. When adding electrons to MnSi by cobalt doping, the helimagnetic structure is conserved with a decrease of the helix pitch up to a concentration $x=0.04$ in \mncosi \cite{Beille1983}. For higher cobalt content, magnetization measurements on the \mncosi solid solutions have not evidenced any magnetic order above the critical concentration $x_c=0.06$ at which the ferromagnetism is suppressed \cite{Motoya1978}.

The aim of this work was to mimic FeSi physical properties with \mncosi that is isoelectronic for $x=0.5$ and very close regarding its structural parameters. The differences between these two materials, evidenced in the present work, confirm the failure of a rigid band picture to consistently describe the electronic structure of B20 monosilicides.

The impressive amount of both experimental and theoretical work done in the past in this field, is far from exhausting the exciting resources of TM monosilicides. The emerging physics of skyrmions, applied to MnSi and \mncosi enthuses the scientific community \cite{Pfleiderer2009}. On the other hand, new regions of the magnetic phase diagram remain fairly obscure and novel ground states of TM-Si have to be unveiled.

In this paper, a complete characterization is presented of transport, magnetic properties as well as neutron diffraction and muon spin relaxation measurements, of the solid solutions \mncosi. We report the discovery of a new spin glass (SG) state. Spin freezing as well as a ``metal-insulator" transition, are attributed to the formation of localized magnetic moment resulting from chemical disorder.

\section{Sample preparation and experimental details}

Polycrystalline samples were synthesized using a home-made arc furnace starting from 4N purity transition metals and 6N silicon chunks, mixed in stoichiometric amount. Annealing at 900$^o$C for 48 hours in high vacuum (about $5\times 10^{-7}$ mbar) is necessary to improve the crystalline quality and chemical homogeneity of the solid solutions. For magnetoresistance (MR) and neutron diffraction measurements, single crystals were grown by the Czochralski pulling from a levitating melt under 3 bar of argon.

X-ray powder diffraction (XRD) was performed in a Philips PW1820 diffractometer using the copper $K_{\alpha}$ radiation ($\lambda=1.5406$ {\AA}). The XRD spectra were analyzed with a full pattern profile refinement method using the Fullprof program suite \cite{Fullprof}.
DC magnetic susceptibility and magnetization data were obtained with a Quantum design Magnetic Property Measurement System (MPMS2) with a squid magnetometer. AC-susceptibility measurements were carried out in a Quantum Design Physical Property Measurement System (PPMS) with an ac excitation field $H = 1$ Oe for a set of four frequencies from 10 Hz to 10 kHz. Electrical magnetoresistance was measured using a standard DC four-probe setup.

Muon-spin spin relaxation experiments ($\mu$SR) were performed on GPS (down to 2~K) and LTF (down to 50~mK) instruments at the Swiss Muon Source (S$\mu$S) and single-crystal neutron-diffraction on TASP (triple-axis spectrometer \cite{Semadeni2001}) instruments at SINQ (both facilities located at the Paul Scherrer Institute, Villigen, Switzerland). The reported $\mu$SR data were obtained in a longitudinal field of 5~mT to quench the depolarization from the $^{55}Mn$ nuclear magnetic moments.

\section{Results and discussion}

\begin{figure}[htp]
  \includegraphics[width=8.5cm]{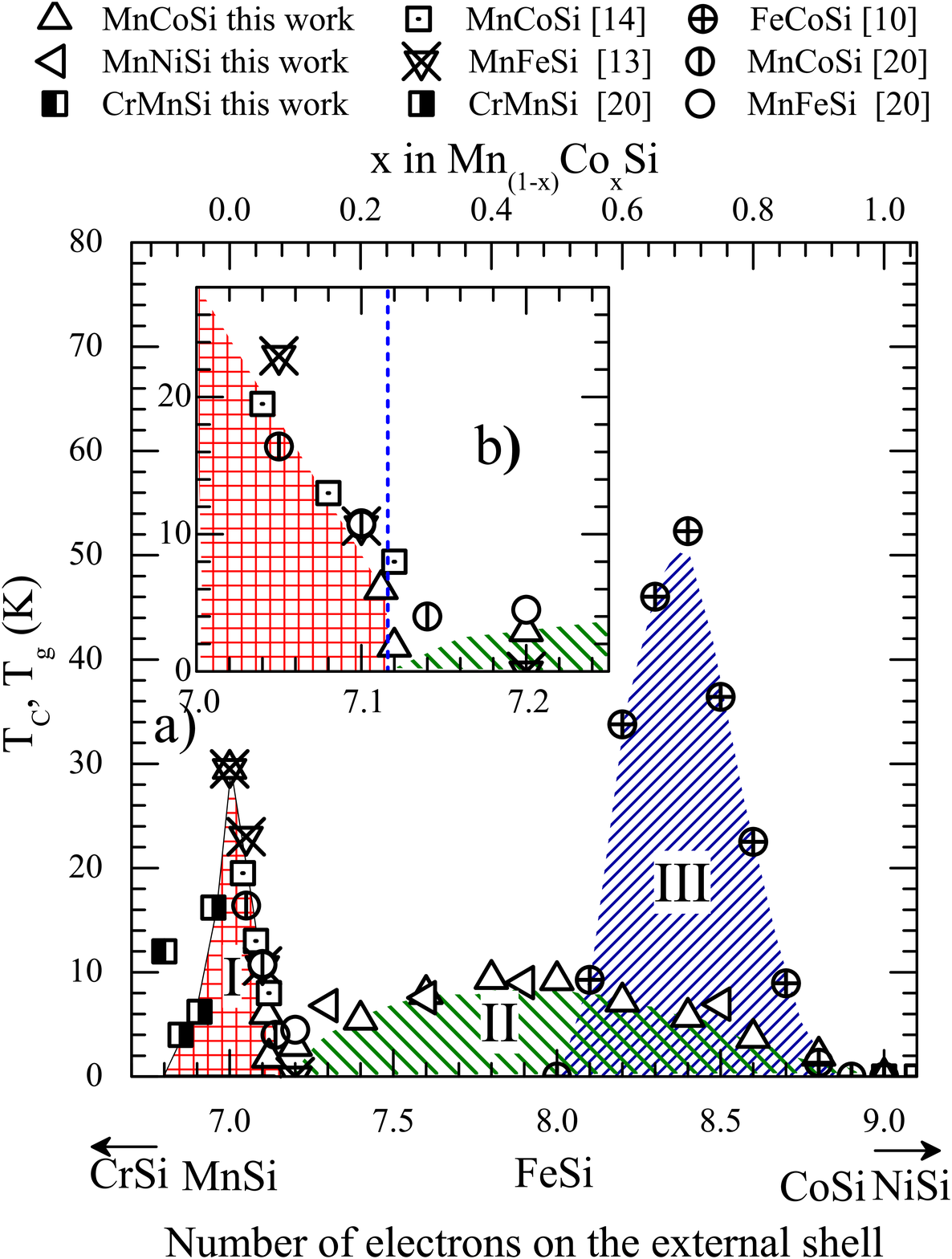}\\
  \caption{(color online) a) Magnetic ordering temperatures of monosilicides. b) Expended scale of the region around $x=0.05$ of the magnetic phase diagram. The dashed vertical line indicates the concentration of the helimagnetic-spin glass transition found in \mncosi in this work.}\label{Tc}
\end{figure}

The magnetic phase diagram is presented in Fig.~\ref{Tc}. The ordering temperature of the various magnetic ground states that have been observed in the family of B20 silicides is plotted as the function of the number of electrons in the external shell.

\begin{figure}[htp]
  \includegraphics[width=8.5cm]{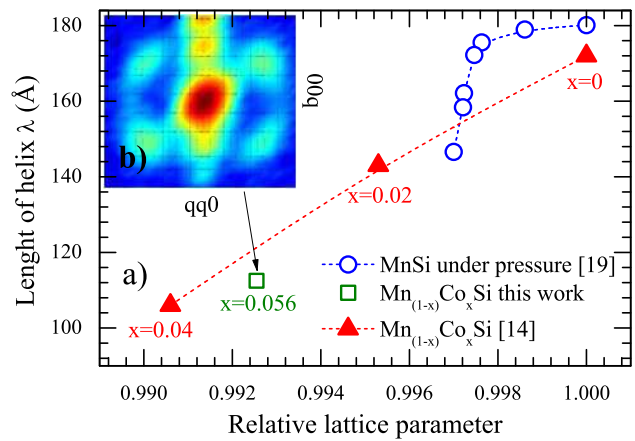}\\
  \caption{(color online) a) Evolution of the length of the helix as a function of the relative lattice parameter in the case of hydrostatic pressure on MnSi \cite{Fak2005} and chemical substitution of Mn with Co \cite{Beille1983}. b) Neutron intensity color map taken at 1.8 K in the vicinity of the (110) nuclear Bragg peak showing the magnetic satellites of the helimagnetic structure along the [111] direction.}\label{pitch}
\end{figure}

In addition to the \fecosi solid solution (region III in Fig.~\ref{Tc}) and pure MnSi, helimagnetism is also observed in a narrow part of the phase diagram close to MnSi (region I in Fig.~\ref{Tc}). The doping-induced decrease of the ordering temperature of MnSi does not depend on the transition metal used as a doping element but only on the number of electrons added to (Co, Ni, Fe) or removed from (Cr) the system. Neutron diffraction reveals magnetic satellites below $T_C$ (Fig.~\ref{pitch} b)) along the [111] direction, indicating that the helimagnetic order is preserved with an helix period that decreases with increasing cobalt content (Fig.~\ref{pitch} a)). Both hydrostatic pressure \cite{Fak2005} and Mn chemical substitution \cite{Beille1983,Achu1998} in MnSi, shrink the lattice and reduce the helix pitch by a comparable amount. Our result for Mn$_{0.944}$Co$_{0.056}$Si with an helix wavelength $\lambda=112$ $\AA$ for a lattice parameter reduction of $0.8\%$ are consistent with previous reports (Fig.~\ref{pitch} a)).

Despite different trends in the two experimental curves ($\lambda=f(P)$ and $\lambda=f(x)$) (maybe due to disorder, electron doping in \mncosi, uncertainty of the applied pressure or intrinsic error bars) the shortening of the helix is the signature of a smooth transition of magnetic exchange from a positive to a negative value that progressively drives the system in a more antiferromagnetic configuration. Because of the very low symmetry of the Mn site and its high coordination (3 different distances of the 7 first neighbors), the exchange is very sensitive to displacive and chemical disorder. Recently, Manyala et al. showed that, by doping MnSi with iron, the Curie-Weiss temperature was switched from positive ($x<0.2$ in \mnfesi) to negative values, suggesting an antiferromagnetic exchange \cite{Manyala2008}.\\

\begin{figure}[htp]
  \includegraphics[width=8.5cm]{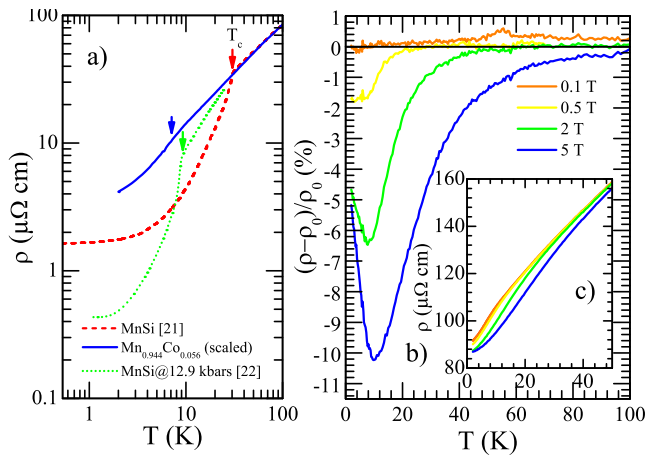}\\
  \caption{(color online)a) Comparison of the temperature dependence of the resistivity of MnSi \cite{Kadowaki1982} (red dashed curve), Mn$_{0.944}$Co$_{0.056}$Si (blue solid curve) and MnSi at $P=12.9$ kbar \cite{Pfleiderer1997} (green dotted curve). b)-c) relative magnetoresistance $\Delta \rho/\rho_0$ and resistivity $\rho(T)$ for Mn$_{0.944}$Co$_{0.056}$Si. Magnetic field values are given in the figure. }\label{MR095}
\end{figure}

In the composition range of region I of Fig.~\ref{Tc}, the temperature dependence of the resistivity evolves in a similar manner as a function of cobalt doping and external pressure. Figure~\ref{MR095} a) shows a comparison of the temperature dependence of the resistivity of Mn$_{0.944}$Co$_{0.056}$Si (blue solid curve) with the one of MnSi \cite{Kadowaki1982} (red dashed curve) and MnSi under an hydrostatic pressure of $12.9$ kbar \cite{Pfleiderer1997} (green dotted curve). Our experimental data have been scaled as the scattering caused by disorder as well as the residual resistivity are slightly larger than those of MnSi.

In the same composition range, a negative magnetoresistance (Fig.~\ref{MR095} b)) with a minimum at $T_c$ was measured, very similar to previous reports on MnSi \cite{Kadowaki1982}.

\begin{figure}[htp]
  \includegraphics[width=8.5cm]{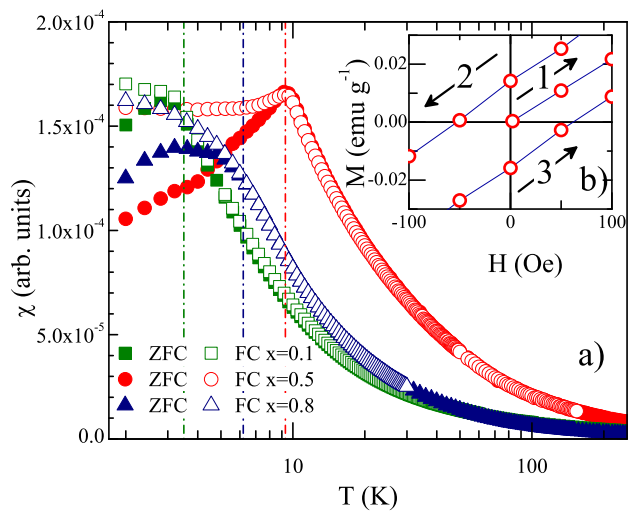}\\
  \caption{(color online) a) Temperature dependence of the zero field cooled (ZFC) (full symbols) and field cooled (FC) (open symbols) DC magnetic susceptibility for three different compositions of \mncosi ($x=0.1$, $x=0.5$ and $x=0.8$). The temperature of the magnetic transition is indicated as a vertical dash-dotted line as the FC and ZFC curves start to overlap. b) Hysteresis loop for the sample Mn$_{0.5}$Co$_{0.5}$Si at 5K. The field sequence is numbered.}\label{suscep}
\end{figure}

When 6 at.\% of cobalt substitution for manganese is performed, the system does not exhibit ferromagnetic ordering down to 100~mK. When the cobalt content is further increased, the magnetic susceptibility exhibits a transition, visible as a peak and marked by a dashed dotted line for three different compositions in Fig.~\ref{suscep} a). The transition temperatures, as reported in Fig.~\ref{Tc}, are defined as the onset of field-cooled/zero-field-cooled irreversibility. At temperatures below the transition, a weak hysteresis loop with a coercitive field of 50~Oe is observed (see Fig.~\ref{suscep} b) and neither satellites nor evidence of long range order were detected using neutron diffraction.

All these experimental results exclude long range magnetic ordering in \mncosi for $x>0.06$, and already hint at the formation of a spin glass (SG) over the whole range of region II in Fig.~\ref{Tc}.

When $\theta_{CW}>0$, the system may exhibit a ferromagnetic state as illustrated by many experimental exemples. When $\theta_{CW}<0$, the system balances between a standard antiferromagnetic long range order and a SG state. Shell et al. noted that SG behavior can be observed as far as $\theta_{CW}\simeq-2.5\times T_g$. Over the whole range of composition, the material exhibits pure Curie-Weiss behavior only for very small cobalt doping and around $x=0.50$, where $\theta_{CW}=-9$ $K$. For this composition, the ratio $\theta_{CW}/T_g \simeq -1$ is close to that reported for disordered antiferromagnetic spin glass (Cu$_3$Pt)$_{(1-x)}$Mn$_x$ \cite{Shell1982}.\\

\begin{figure}[htp]
  \includegraphics[width=8.5cm]{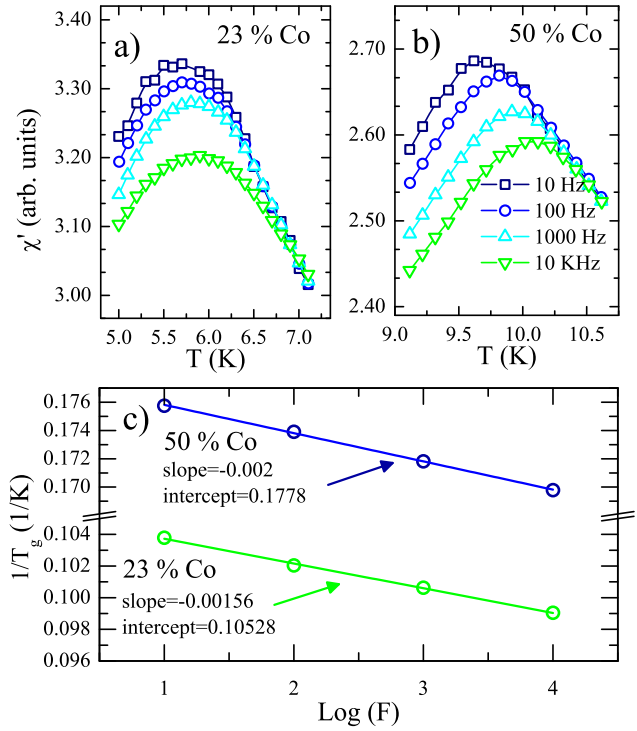}\\
  \caption{(color online) In-phase component of the AC magnetic susceptibility for a) Mn$_{0.77}$Co$_{0.23}$Si and b) Mn$_{0.5}$Co$_{0.5}$Si at 4 different frequencies. c) Frequency dependence of the inverse of the freezing temperature $1/T_g$ versus $log(F)$. The parameters of linear fits (dashed lines) are reported on the figure for each composition.}\label{khi_f}
\end{figure}

The freezing of magnetic moments below the SG transition temperature $T_g$ implies a frequency dependent peak in the AC susceptibility. Figure~\ref{khi_f} shows the temperature dependence of the AC magnetic susceptibilities for $x=0.23$ and $x=0.50$ in \mncosi.

The temperature dependence of the in-phase component ($\chi '$) obtained at the lowest frequency (10~Hz) shows a peak at $T_g=5.73$ $K$ and $T_g=9.65$ $K$ for $x = 0.23$ and $x=0.50$, respectively. The inverse of $T_g$ varies linearly with the log of the frequency as typically reported for metallic spin glasses \cite{Binder1986} (Fig.~\ref{khi_f} c)). Tholence proposed that the ratio $\Delta(T_g)/\Delta(Log(F))$ is proportional to the concentration $x$ of the dopant in CuMn and AgMn spin glasses \cite{Tholence1981}. In the table \ref{param}, we show that this scenario is supported by our experimental observation and the the values of the slopes are consistent with previous reports.\\
\begin{table}
  \centering
  \begin{tabular}{l|c|c}
    \hline\hline
    & $x=0.23$ & $x=0.50$\\
    \hline
    $\Delta T_g$& 0.2 & 0.461\\
    $\Delta log(F)$&3&3\\
    $\Delta T_g/\Delta log(F)$&0.0667&0.1537\\
    $\Delta T_g/\Delta log(F)\cdot x^{-1}$&0.290&0.307\\
    \hline\hline
  \end{tabular}
  \caption{parameters of the dependence of $\Delta T_g/\Delta log(F)$ with the cobalt doping in the spin glass.}\label{param}
\end{table}

The SG behavior is also supported by muon spin relaxation experiments.

\begin{figure}[btp]
  \includegraphics[width=8.5cm]{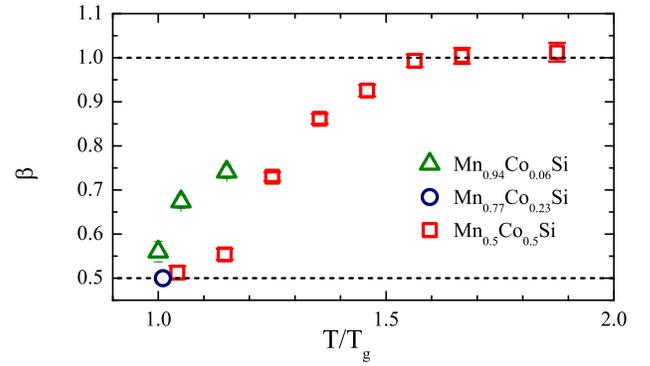}\\
  \caption{(color online) temperature dependence of the power $\beta$ in eq.~\ref{campbell} for \mncosi samples with $x=0.06$, $x=0.23$ and $x=0.50$ on a reduced temperature scale. The value goes from $\beta=1$ above $T_g$ (paramagnetism) to $\beta=\frac{1}{2}$ at $T_g$ upon cooling.} \label{beta}
\end{figure}

In a spin glass systems above its freezing temperature, it was shown that the muon polarisation decay can be expressed by a stretched exponential function of the form:
\begin{equation}
G_z(t)=\exp(-(\lambda_d t)^{\beta})
\label{campbell}
\end{equation}
where $\lambda_d$ is the muon depolarization rate. For temperatures much higher than $T_g$ ($T\geq4\times T_g$), i.e. when the system is in a conventional paramagnetic state, one observes $\beta=1$, indicating that the muon senses rapidly fluctuating fields. Upon decreasing the temperature towards $T_g$ one observes $\beta<1$. Two limits are usually discussed, corresponding either: i) to a situation where the coupling between the muons and the local spins has a given distribution, but the local spins have a unique relaxation time at each temperature (in this case one obtains $\beta = \frac{1}{2}$ \cite{Uemura1984}); or ii) to the so-called concentrated limit, where one assumes an (ideally) unique value for the coupling constant, but a distribution of local spins relaxation time (here a limit $\beta=\frac{1}{3}$ is observed \cite{Campbell1994}). For our systems, the temperature dependence of $\beta$ for $x=0.06$ and $x=0.5$ is plotted as a function of the reduced temperature $T/T_g$ in Fig.~\ref{beta}.

We clearly see the limit $\beta=\frac{1}{2}$ at $T_g$ and the value of $\beta=1$ reached at a temperature $T\simeq 2\times T_g$. The evaluation of the temperature evolution of the $\beta$ parameter was only possible for $x=0.50$ and for a few temperatures above $T_g$ for $x=0.06$ samples where the $\lambda$ value is sufficiently large to keep the fit reliable. Nevertheless, for $x=0.23$ the value of $\beta$ just above $T_g$ is $0.5$. This experimental fact points to a rather wide distribution of local magnetic environments for the implanted muons, which is online with the rather high substitution of manganese by cobalt in our systems.

\begin{figure}[btp]
  \includegraphics[width=8.5cm]{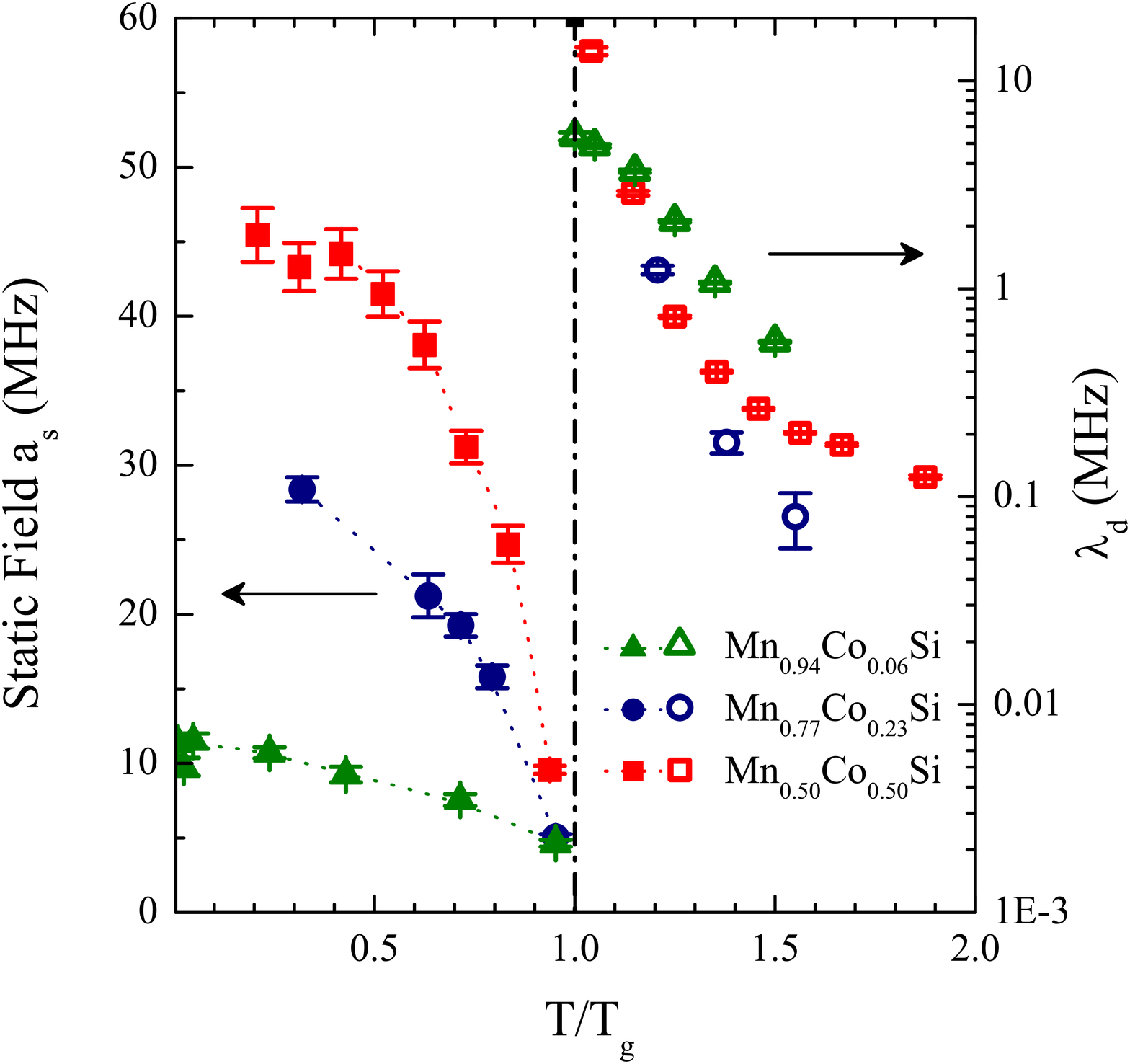}\\
  \caption{(color online) (Left axis): temperature dependence of the static field ($a_s$) below the spin glass transition $T_g$ from the fit of muon spin relaxation signal with Eq.~\ref{uemura_eq} for cobalt concentrations $x=0.06$ (green symbols), $x=0.23$ (blue symbols) and $x=0.50$ (red symbols). $T_g$ is indicated as vertical dash-dotted lines. (Right axis): temperature evolution above $T_g$ of the depolarization rate ($\lambda_d$) from Eq.~\ref{campbell}.}\label{muon}
\end{figure}

When discussing the $\mu$SR measurements below $T_g$, the observation of a $\beta=\frac{1}{2}$ parameter leads us to naturally assume the analytical expression for the muon spin relaxation function $G_z(t)$ proposed by Uemura et al. \cite{Uemura1984}:
\begin{eqnarray}
\label{uemura_eq}\nonumber G_z(t)&=&\frac{1}{3}\exp(-\sqrt{\lambda_d t})+\frac{2}{3}\Big(1-\frac{a_s^2t^2}{(\lambda_d t+a_s^2 t^2)^\frac{1}{2}}\Big)\\
&& \times \exp(-\sqrt{\lambda_d t+a_s^2t^2})\\
\lambda_d&=&\frac{4a_d^2}{\nu}\label{depol_eqn}\\
Q&=&\frac{a_s^2}{a_s^2+a_d^2}
\end{eqnarray}
where $a_s$ and $a_d$ represent respectively the average amplitudes of the static and dynamic random local field at the muon site. The full field $a$ at the muon site can be expressed as:
\begin{equation}
a=\sqrt{a_s^2+a_d^2}~.
\end{equation}
Note that the Eq.~\ref{campbell} with $\beta=\frac{1}{2}$ represents the limiting case of Eq.~\ref{uemura_eq} when $a_s = 0$.
The parameter $a_s$ exhibits nonzero values only below $T_g$, i.e. signaling the occurrence of a static field at the muon site. The temperature evolution of the static field below $T_g$ for 3 different compositions ($x=0.06$, $x=0.23$ and $x=0.50$ in \mncosi) is shown in the left panel of the Fig.~\ref{muon}. Note also that the fits provide values of $a_d$ rather small for each cases. It is worthwhile to note that the limit of Eq.~\ref{uemura_eq} when $a_d = 0$ is the so-called Lorentz Kubo-Toyabe function \cite{Kubo1981}, which is valid in the case of strongly disordered static magnetism. A posteriori, this observation constitutes an additional argument for the validity of the wide distribution of coupling between muons and local spins observed in the dynamical regime.

With the knowledge acquired in the static regime, we discuss now the temperature dependence of the muon depolarization rate in the dynamical regime, which diverges when approaching $T_g$ from higher temperatures due to critical fluctuations \cite{Malureanu2003}. The correlation time $\tau_c=1/\nu$ of the magnetic moments at the muon site can be deduced from $\lambda_d$ using Eq.~\ref{depol_eqn} and assuming that the value of the fluctuating field above $T_g$ correspond to the extrapolated limit of the static field in the SG state (in other words $a_d(T>T_g) = a_s(T\rightarrow 0)$;
corresponding to $a_d=12$~MHz for $x=0.06$, $a_d=32$~MHz for $x=0.23$ and $a_d=48$~MHz for $x=0.5$). Assuming also that $\tau_c=\tau_0[T/(T-T_g)]^2$ \cite{Uemura1981}, the temperature evolution of $\lambda_d$  follows:
\begin{equation}\label{tau}
\lambda_d = 4a_d^2\tau_0\Big(\frac{T}{T-T_g}\Big)^2
\end{equation}
with $\tau_0=3.77\cdot10^{-18}$ $s$ for $x=0.23$ and $\tau_0=3.81\cdot10^{-18}$ $s$ for $x=0.5$. It is visible on Fig.~\ref{muon} that $\lambda_d$ does not exhibit a divergence at $T_g$ for $x=0.06$, concentration close to the limit of existence of the spin glass state.\\

\begin{figure}[htp]
  \includegraphics[width=8.5cm]{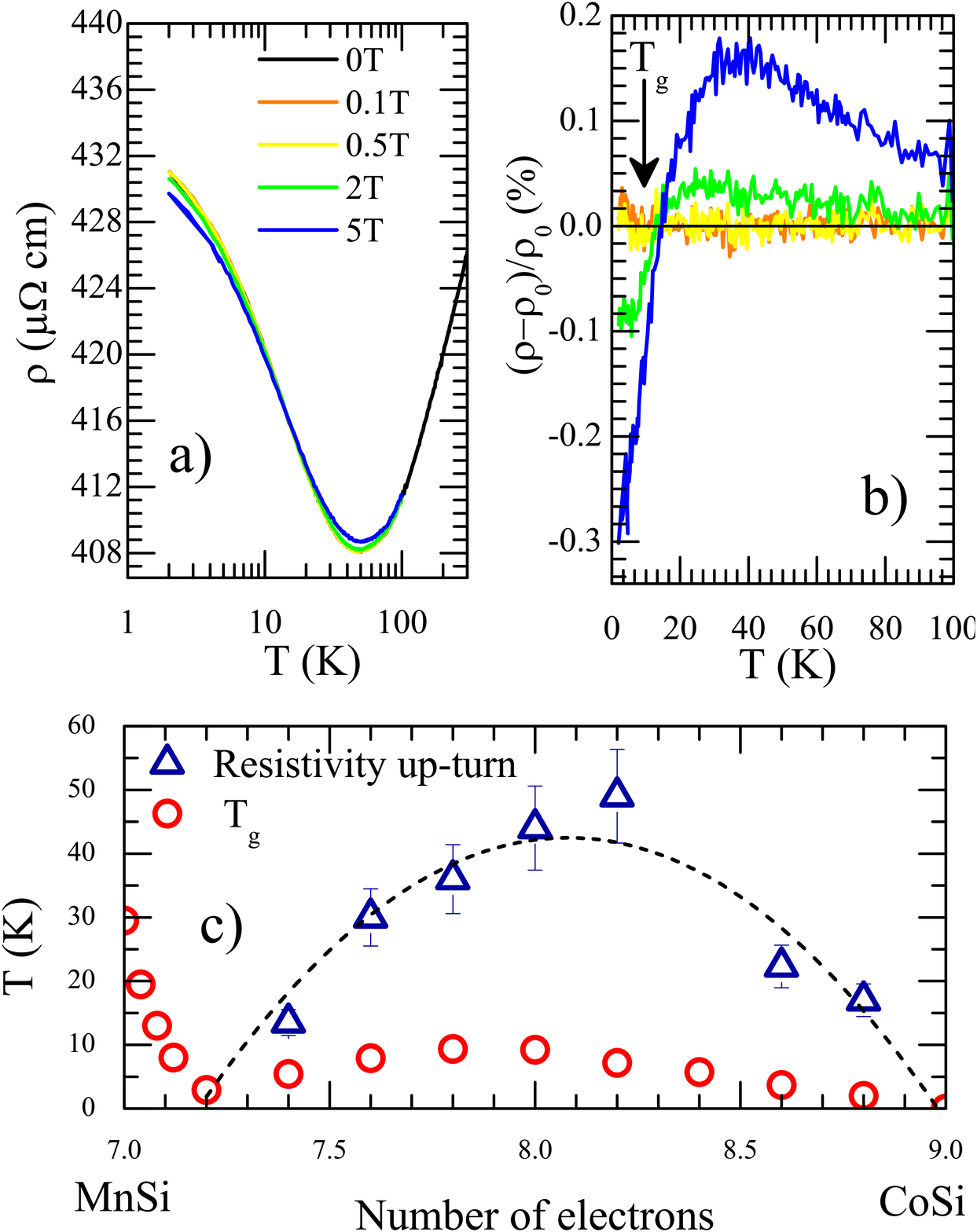}\\
  \caption{(color online) Temperature dependence of a) the resistivity $\rho$ b) the relative magnetoresistance $\Delta \rho/\rho_0$ for Mn$_{0.5}$Co$_{0.5}$Si. Magnetic fields are given in the figure. c) Temperature dependence of $T_g$ and the temperature of the upturn of resistivity. The dashed line corresponds to the formula of magnetic ``disorder resistivity" defined in the text \cite{Dekker1962}.}\label{MR05}
\end{figure}

In the composition range of the SG state, the resistivity exhibits a non monotonic temperature dependence (Fig.~\ref{MR05} a)) with an minimum that scales with the freezing temperature $T_g$ as $T_{\textrm{upturn}}\simeq 2.4\times T_g$) (Fig.~\ref{MR05} c)).

In magnetic systems with antiferromagnetic exchange, such a ``metal-insulator" transition could be ascribed to the onset of the Kondo effect. The absolute value of the magnetoresistance is very small (less than 1\%) and presents two distinct behaviors. At high temperature, a positive magnetoresistance could be attributed to standard Kohler contribution or quantum interference effects as pointed out by Manyala et al. on \fecosi \cite{Manyala2000}. At low temperature, the magnetoresistance exhibits a crossover due to a negative contribution as usually seen in Kondo systems \cite{Teyssier2009}.

In the scheme of a spin-$\frac{1}{2}$ Kondo model, the magnetoresistance should be much larger than what is observed here. Indeed, a modest field usually suppresses the Kondo effect and should give rise to a magnetoresistance of about $65\%$. Moreover, a pair (or a cluster) of magnetic impurities very close to each other, will interact in such a way that the pair (or the cluster), viewed as a single entity, will have a lower Kondo temperature than a single impurity \cite{Hertz1979}. In a highly concentrated alloys like \mncosi, the number of magnetic clusters increases with doping thus lowering the effective Kondo temperature. The freezing of the effective moments can then occur only when the $T_K$ is driven to zero. This is in contradiction with the fact that the upturn in resistivity is not affected by the spin glass formation, and that its maximum temperature is obtained at large doping.

Although we cannot completely exclude Kondo effect in small volumes, where it is favored by a random chemical environment of the transition metal, it cannot be responsible for the upward deviation of the resistivity.

A minimum in the resistivity necessarily requires a temperature dependent scattering rate. In the early 60's, Dekker \cite{Dekker1965} proposed a model for the resistivity in a binary alloy. The magnetic disorder due to the random distribution of two atoms (A and B) with different electronic structures adds a concentration dependent term in the expression of the resistivity. The sign of this correction depends on the sign of the exchange interaction (positive or negative for AFM or FM exchange, respectively) \cite{Dekker1965, Dekker1962}. This ``disorder resistivity", proportional to $x(1-x)$ (where $x$ and $(1-x)$ are the respective concentrations of A and B ions), has a maximum for $x=0.5$. Indeed, the enhancement of the resistivity due to this extra term has been found to be maximum in \mncosi for $x=0.55$. The composition dependent $T_{\textrm{upturn}}$ varies accordingly (dashed curve in Fig.~\ref{MR05} c). The fact that the maximum is slightly shifted away from $x=0.5$ probably reveals the asymmetry of the density of states around the Fermi level. \\

The scattering centers described above could be viewed as virtual bound states undergoing spin fluctuations (Localized Spin Fluctuations, LSF) as proposed by Rivier and Zlatic \cite{Rivier1972}. Based on the Anderson formalism \cite{Anderson1961}, this model was adopted to explain resistivity minima in spin glasses like PdCr \cite{Cochrane1979},RhCo \cite{Jamieson1975}, RhFe \cite{Rusby1974}, (V,Cr)Fe \cite{Rusby1974b}, AuV and AlMn \cite{Rizzuto1973} and more generally in alloys made of atoms with different magnetic moments \cite{Coles1974}.\\

\begin{figure}[htp]
  \includegraphics[width=8.5cm]{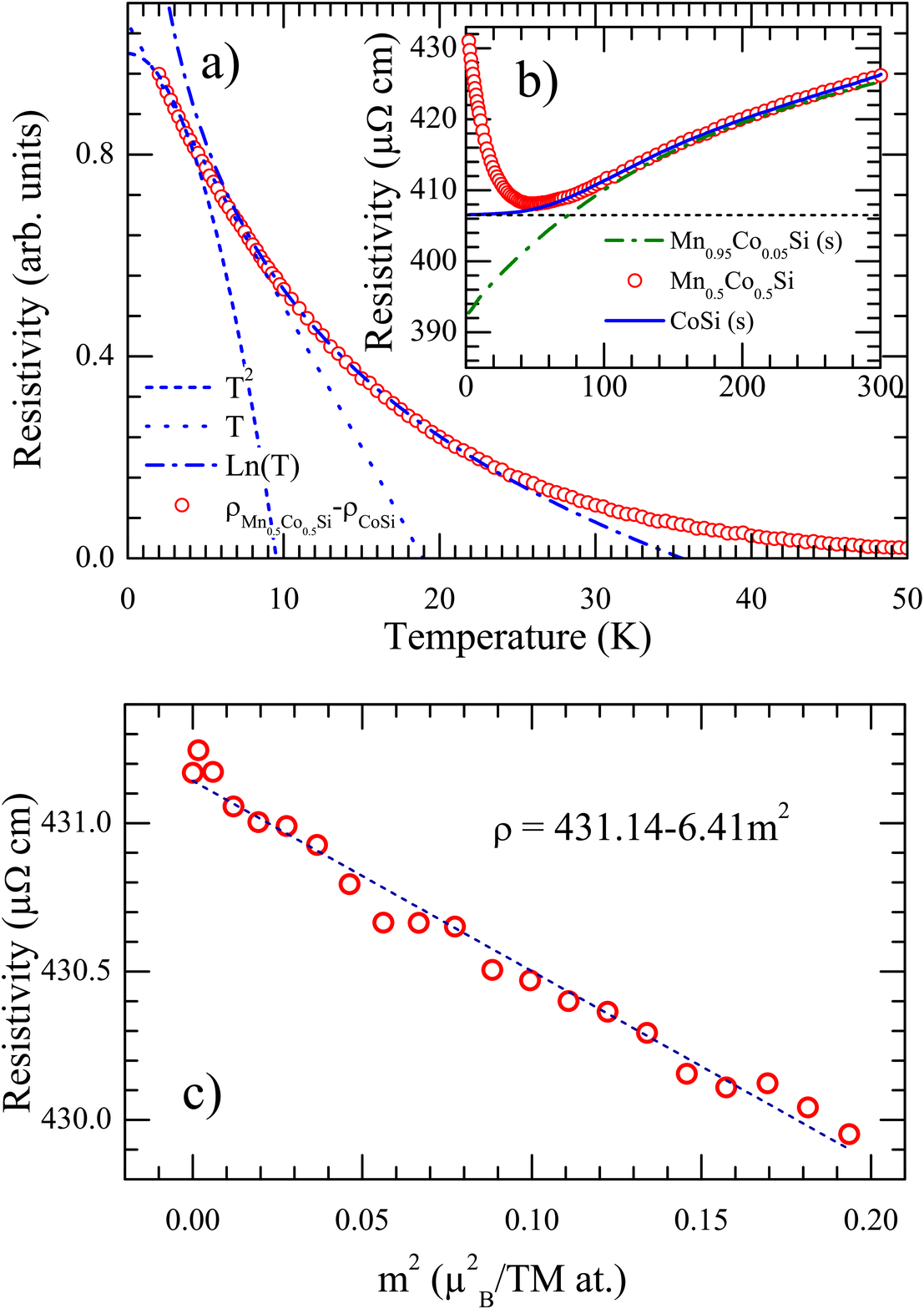}\\
  \caption{a) Temperature dependence of the resistivity of Mn$_{0.5}$Co$_{0.5}$Si. The dotted lines represent the three regions, parabolic (fast LSF), linear and logarithmic regimes (slow LSF) \cite{Rivier1972}. The contribution of phonons is removed by substraction of the scaled (s) resistivity of CoSi (b). The scaled resistivity curve of Mn$_{0.95}$Co$_{0.05}$Si is shown as a dash-dotted line. c) Resistivity versus the square of the magnetic moment per transition metal atomic site at $T=5$~K. }\label{rivier}
\end{figure}

According to this model, the resistivity has a finite limit$\rho_0$ at $T=0$ and decreases with increasing temperature. The behavior at low temperature is $\rho=\rho_0-AT^2$ followed by three crossovers passing trough $-T$, $1-\ln(T)$ and $T^{-1}$. The high temperature $\rho(T)$ for Mn$_{0.94}$Co$_{0.06}$Si, Mn$_{0.5}$Co$_{0.5}$Si, CoSi overlap after scaling, indication of a similar phonon contribution over the whole composition range (Fig.~\ref{rivier} b)). The temperature dependence of the ``electronic resistivity" of Mn$_{0.5}$Co$_{0.5}$Si, obtained by subtracting the scaled curve of CoSi, is shown in Fig.~\ref{rivier} a). The $T^2$, $T$ and $\ln(T)$ curves clearly reveal the good agreement of our experimental data with the model over a wide temperature range. The lack of data below 2K makes the parabolic temperature dependence ill defined. However, Fig.~\ref{MR05} a) shows that the resistivity tends to a plateau with a finite value of the resistivity at zero temperature of about $\rho_0\simeq 435$ $\mu \Omega$ cm. This agrees with the extrapolation to $T=0$ of the parabola of Fig.~\ref{rivier} a).\\

When cobalt substitutes for manganese in \mncosi, electrons are added to the system (2$e^-$ per formula unit) and deeper potential wells are randomly distributed on the transition metal sublattice. Upon cooling, this potential grid can trap electrons and because of the Hund rules and Coulomb repulsion, a local magnetic moment is formed. A similar approach was chosen by J. Mathon, who proposed a Hubbard based model to describe the creation of local magnetic moments and the formation of the spin glass state in an itinerant isoelectronic alloy, in the presence of an antiferromagnetic exchange \cite{Mathon1978}.\\

Applying a magnetic field to the system tends to decrease the magnetic disorder, reducing its contribution to scattering, thus resulting in a negative magnetoresistance (Fig.~\ref{MR05} b)). The complete surpression of the ``magnetic disorder resistivity" term should be achieved when the magnetic moments are frozen along the applied field direction. In this situations, the absence of magnetic fluctuations should give a magnetization corresponding to that expected for the Curie-Weiss effective moment. Since the scattering does not depend on the sign of the magnetic moment, the simplest $m$ dependence is quadratic. Thus Fig.~\ref{rivier} c) shows a plot of the resistivity versus the square of the magnetic moment per transition metal atomic site, measured at $5$ K. The fit equation is given in the figure. According to this expression, the value of the magnetic moment that would suppress magnetic disorder ($\rho=\rho_{res}$ as defined in Fig.~\ref{rivier} b) is $m\simeq 1.97$ $\mu_B$). This value is close to the Curie-Weiss value for Mn$_{0.5}$Co$_{0.5}$Si ($\mu=1.78$ $\mu_B$).

One goal of this study was to establish similarities in electronic structures of isoelectronic TM monosilicides. In this picture, Mn$_{0.5}$Co$_{0.5}$Si was supposed to be equivalent to FeSi and thus to exhibit similar semiconducting behaviors. From an unified model explaining the formation of local moment in MnSi and FeSi \cite{Evangelou1983}, the origin of the gap in FeSi could be viewed as an ultimate signature of the localization effects described above \cite{Schlesinger1993}.

\section{Conclusion}

We report the presence of a spin glass state in \mncosi for a wide composition range from $x=0.05$ to $x=1$. This new ground state is the first example of an antiferromagnetic-like order in a magnetic transition metal monosilicides. Upon cooling, randomly distributed localized magnetic moments form due to chemical disorder. They scatter the remaining conduction electrons, resulting in an upturn in the resistivity. When the concentration of localized magnetic moment is sufficiently large, the SG forms. The discovery of such a new ground state shows the high interest in studying ternary solid solutions of transition metal monosilicides.

\section*{Acknowledgments}

The authors thank J.~Ditusa and C.~Pfleiderer for sharing their expertise on the B20 silicides. Thanks to R.~Lortz, B. Roessli and A.~Piriou for their help in physical properties measurements.
This work is supported by the Swiss National Science Foundation through grant 200020-109588 and the National Center of Competence in Research (NCCR) ``Materials with Novel Electronic Properties-MaNEP". Part of this work was performed at the Swiss Muon Source (S$\mu$S) and the Swiss Spallation Neutron Source (SINQ) of the Paul Scherrer Institute, Villigen, Switzerland.

\end{document}